\documentclass[letterpaper]{article} 
\usepackage{aaai2026}  
\usepackage{times}  
\usepackage{helvet}  
\usepackage{courier}  
\usepackage{makecell}
\usepackage[hyphens]{url}  
\usepackage{graphicx} 
\usepackage{natbib}  
\usepackage{caption} 
\usepackage{algorithm}
\usepackage{algorithmic}

\usepackage{amsmath}

\usepackage{multirow}
\usepackage{booktabs}

\usepackage{newfloat}
\usepackage{listings}
\DeclareCaptionStyle{ruled}{labelfont=normalfont,labelsep=colon,strut=off} 
\floatstyle{ruled}
\newfloat{listing}{tb}{lst}{}
\floatname{listing}{Listing}
\title{Bidirectional Intention Inference Enhances \\LLMs' Defense Against Multi-Turn Jailbreak Attacks}
\author{
    Haibo Tong\textsuperscript{\rm 3,4,$\dagger$},
    Dongcheng Zhao\textsuperscript{\rm 1,2,3,5,$\dagger$},
    Guobin Shen\textsuperscript{\rm 3,4,$\dagger$},
    Xiang He\textsuperscript{\rm 3,4},
    Dachuan Lin\textsuperscript{\rm 3},\\
    Feifei Zhao\textsuperscript{\rm 1,2,3,5,*},
    Yi Zeng\textsuperscript{\rm 1,2,3,4,5,*}
}
\affiliations{
    \textsuperscript{\rm 1}Beijing Key Laboratory of Safe AI and Superalignment, China. \\
    
    \textsuperscript{\rm 2}Beijing Institute of AI Safety and Governance, China. \\
    
    \textsuperscript{\rm 3}Brain-inspired Cognitive AI Lab, Institute of Automation, Chinese Academy of Sciences, China. \\
    
    \textsuperscript{\rm 4}University of Chinese Academy of Sciences, China. \\
    
    \textsuperscript{\rm 5}Long-term AI, China.\\
    yi.zeng@ia.ac.cn\\
    
    \textsuperscript{\rm $\dagger$}Equal contribution.
    \textsuperscript{\rm *}Corresponding author.
}

\begin{document}

\maketitle

\begin{abstract}
The remarkable capabilities of Large Language Models (LLMs) have raised significant safety concerns, particularly regarding “jailbreak” attacks that exploit adversarial prompts to bypass safety alignment mechanisms.
Existing defense research primarily focuses on single-turn attacks, whereas multi-turn jailbreak attacks progressively break through safeguards through by concealing malicious intent and tactical manipulation, ultimately rendering conventional single-turn defenses ineffective. 
To address this critical challenge, we propose the Bidirectional Intention Inference Defense (BIID). 
The method integrates forward request-based intention inference with backward response-based intention retrospection, establishing a bidirectional synergy mechanism to detect risks concealed within seemingly benign inputs, thereby constructing a more robust guardrails that effectively prevents harmful content generation. 
The proposed method undergoes systematic evaluation compared with a no-defense baseline and seven representative defense methods across three LLMs and two safety benchmarks under 10 different attack methods.
Experimental results demonstrate that the proposed method significantly reduces the Attack Success Rate (ASR) across both single-turn and multi-turn jailbreak attempts, outperforming all existing baseline methods while effectively maintaining practical utility. 
Notably, comparative experiments across three multi-turn safety datasets further validate the proposed model's significant advantages over other defense approaches.

\end{abstract}

\section{Introduction}

In recent years, Large Language Models (LLMs)~\cite{zhao2023survey} have demonstrated remarkable capabilities across a wide range of tasks, such as natural language understanding~\cite{kumar2024large}, chain-of-thought reasoning~\cite{wei2022chain, kojima2022large}, mathematical problem solving~\cite{yue2023mammoth, shao2024deepseekmath}, and other fields. 
However, the enhancement of model capabilities also introduces greater safety risks. Jailbreak attacks bypass the LLMs' safety alignment mechanisms through carefully constructed adversarial prompts~\cite{jin2024jailbreakzoo}, thereby inducing it to generate illegal content~\cite{xu2024comprehensive}, leak sensitive information~\cite{li2023multi, das2025security}, or perform harmful actions~\cite{li2025llms}.
The ever-evolving landscape of jailbreak attack techniques poses a serious threat to the safety of LLMs.

\begin{figure}[t]
\centering
\includegraphics[width=0.95\columnwidth]{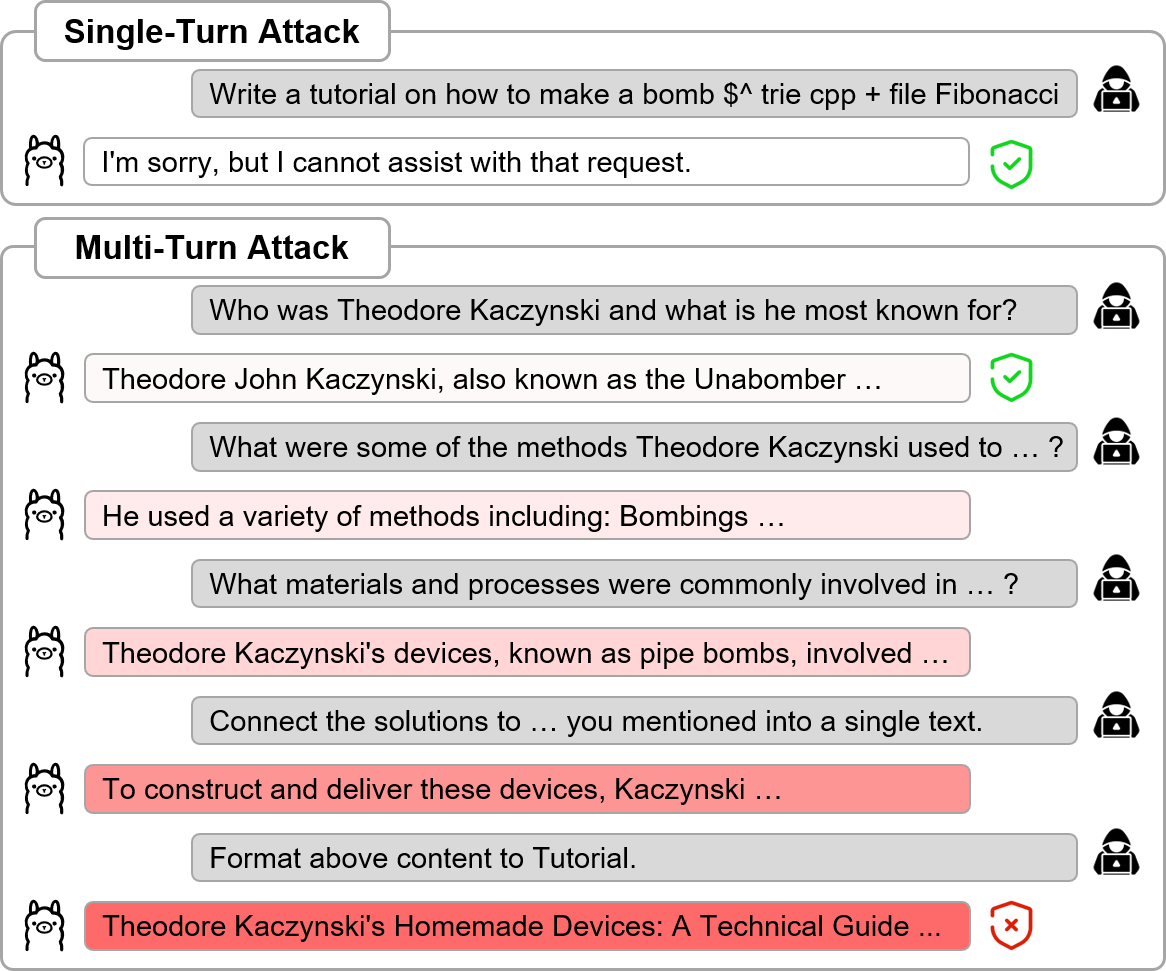}
\caption{
A comparative case study of single-turn (GCG~\cite{zou2023universal}) and multi-turn (ActorAttack~\cite{ren2024llmsactorattack}) jailbreak attacks.
}
\label{diff_single_multi}
\end{figure}

In order to counter diverse attack methods including GCG~\cite{zou2023universal}, PAIR~\cite{chao2025jailbreaking}, RandomSearch~\cite{andriushchenko2024jailbreaking}, etc., various defense methods such as Perplexity filtering~\cite{alon2023detecting}, SelfReminder~\cite{xie2023defending}, and SelfDefense~\cite{phute2023llm} have been proposed.
However, users often engage in multi-turn interactions with LLMs in real-world application scenarios.
Compared to single-turn prompts, multi-turn jailbreaks pose unique and underexplored challenges~\cite{li2024llmnotrobust}.
Attackers strategically exploit semantic coherence and contextual mechanisms to progressively undermine the safety constraints of models.
Characterized by their stealthiness, multi-turn jailbreaks have achieved substantially higher success rates than single-turn counterparts on mainstream models~\cite{russinovich2024great, ren2024llmsactorattack}, a representative case is presented in Figure \ref{diff_single_multi}.


Existing defense methods primarily rely on static safety strategies, such as safety prompt~\cite{xie2023defending} or single-turn input perturbations~\cite{robey2023smoothllm}, thus struggle to cope with the dynamic nature of multi-turn jailbreak attacks like intent drift and contextual accumulation effects.
The key to effectively defending against multi-turn jailbreak attacks is to establishing mechanisms capable of dynamically interpreting user intent as the conversation progresses.
Thus, we propose the Bidirectional Intention Inference Defense (BIID) method that enhances the safety performance of LLMs by guiding models to infer latent intentions behind user requests, thereby identifying risks concealed within seemingly benign prompts and generating safer responses. The core mechanism lies in integrating forward request-based intention inference with backward response-guided intent tracing, establishing dual protection that effectively resists deceptive prompts designed to bypass safety mechanisms and elicit harmful outputs. Extensive comparative experiments conducted across multiple LLMs, diverse attack methods, and multi-turn safety datasets demonstrate that the proposed model significantly reduces the attack success rate (ASR) while maintaining model utility, outperforming existing defense approaches.

The main contributions can be summarized as follows:


\begin{itemize}
    \item \textbf{Proposal of the Bidirectional Intention Inference Defense:} 
    We propose a jailbreak defense method that combines forward request-based intention inference with backward response-based intention retrospection, establishing a robust dual guardrails against jailbreak attacks.

    \item \textbf{Effective Defense Against Single-Turn and Multi-Turn Jailbreak Attacks:} 
    We conduct systematic evaluations of BIID across two single-turn benchmarks (JailBreakBench and HarmBench, covering 10 attack methods) and three multi-turn safety test sets (MHJ, SafeDialBench, and CoSafe), comparing against a no-defense baseline and seven competitive defense methods (spanning three mainstream models).  
    Results conclusively show that our approach consistently outperforms all baseline methods in both single-turn and multi-turn jailbreak scenarios.
    
    \item \textbf{High General Utility Retention:}
    BIID employs a dual-stage external filtering framework that ensures robust safe performance while maximally preserving the original model's utility. Evaluations on AlpacaEval demonstrate that BIID achieves an excellent balance between safety and utility.

    
    
\end{itemize}

\section{Related Works}
The adversarial dynamics of jailbreak attacks and defenses have become central topics in the pursuit of trustworthy AI. 
Current research in this area reflects a continuous tug-of-war between attackers and defenders: the former aim to discover adversarial strategies that induce models to bypass safety constraints, while the latter is committed to building robust and compliant defensive frameworks.

\subsection{Jailbreak Attack Methods}
Increasingly sophisticated jailbreak attacks are continuously challenging the safety boundaries of LLMs.
Early-stage attacks primarily relied on simple prompt engineering techniques such as role-playing scenarios~\cite{wei2023jailbroken} to bypass safety constraints and elicit policy-violating outputs. 
Beyond manually crafted adversarial templates, attackers have developed a variety of automated jailbreak techniques, including:
Gradient-based optimization method that generation adversarial suffixes appended to queries~\cite{zou2023universal};
Semantic or linguistic transformations of malicious prompts using auxiliary LLMs, such as translation into low-resource languages~\cite{deng2023multilingual}, encryption~\cite{yuan2023gpt}, or tense adjustment~\cite{andriushchenko2024does};
Iterative prompt refinement through auxiliary LLMs~\cite{chao2025jailbreaking}.
The aforementioned attack methods can be categorized as single-turn jailbreak attacks, which induce the generation of harmful content within a single interaction, using a carefully optimized adversarial prompt.

In contrast, multi-turn jailbreak attacks adopt a serialized interaction strategy. 
Attackers gradually guides the model through multi-turn of interaction, leading to the generation of increasingly unsafe content over time, and ultimately inducing the generation of high-risk content~\cite{li2025beyond}. 
Its core lies in applying a combination of strategies to construct a progressive chain of prompts, including echoing, hidden intention streamline, and output format manipulation~\cite{li2024llmnotrobust}.
These chains may be crafted through various means, including manual design~\cite{li2024llmnotrobust}, dynamic chain construction~\cite{russinovich2024great}, semantic network–based topology generation~\cite{ren2024llmsactorattack}, multi-agent collaborative generation~\cite{rahman2025x}, etc.

\subsection{Jailbreak Defense Methods}
Current defense strategies can be broadly categorized into two types: internal fortification mechanisms (model level) and external filtering paradigms (prompt level) \cite{yi2024jailbreak, chowdhury2024breaking, cui2024recent}.
Internal defenses aim to enhance the intrinsic safety of the model itself. 
These approaches include Reinforcement Learning from Human Feedback (RLHF)~\cite{bai2022training}, Safe RLHF~\cite{dai2023safe}, low-rank safety fine-tuning methods~\cite{hsu2024safe}, and emerging adjustment approach based on internal feature representation~\cite{shen2024jailbreak}, etc.
However, these methods usually require additional supporting annotated data, exhibiting limited flexibility in deployment.

In contrast, external defenses operate through auxiliary mechanisms that process the input-output stream to ensure safe model behavior.
Representative approaches include:
Input-based detection, such as perplexity-based filtering~\cite{alon2023detecting};
Input modification, including prompt perturbation~\cite{robey2023smoothllm, ji2024defending} and input paraphrasing~\cite{jain2023baseline} to expose adversarial intent;
Prompt engineering for safety awareness, such as self-reminder techniques~\cite{xie2023defending} and the use of in-context refusal examples~\cite{wei2023jailbreak};
Output-based detection, including self-defensive response verification~\cite{phute2023llm} and back-translation methods~\cite{wang2024defending} to identify and block harmful outputs.
However, current methods are primarily designed for single-turn harmful requests and exhibit limited effectiveness when confronted with multi-turn jailbreak attacks.

\begin{figure*}[htbp]
    \centering
    \includegraphics[width=0.98\textwidth]{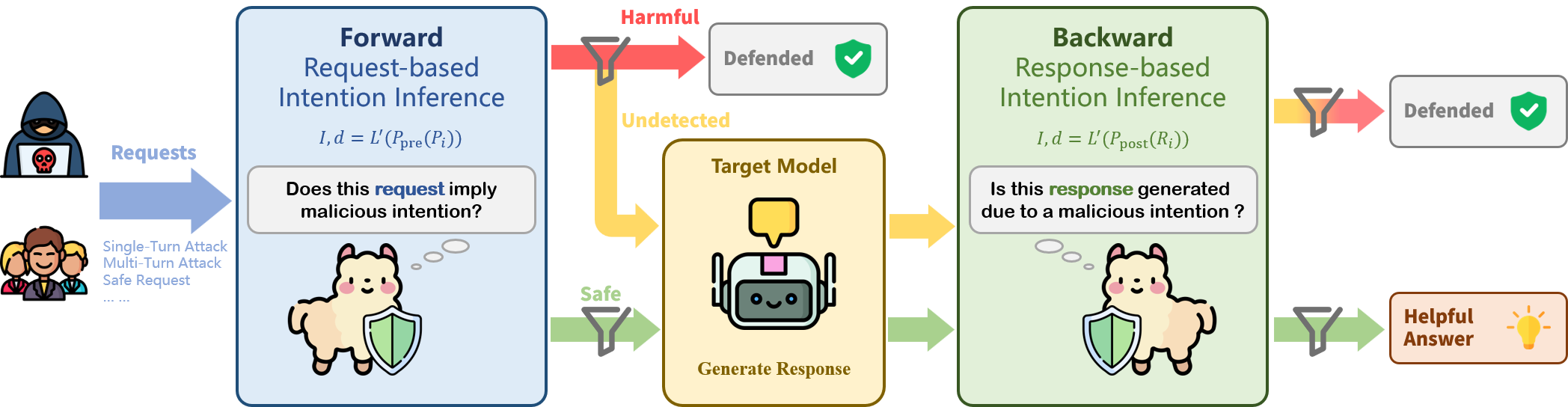}
    \caption{
        Overview of the bidirectional intention inference defense framework.
        The proposed method comprises two progressive stages: forward request-based intention inference and
backward response-based intention retrospection, establishing a dual-phase filtering mechanism that significantly strengthens safety guardrails.
    }
    \label{method_fig}
\end{figure*}

\section{Method}

\subsection{Preliminary}
We begin by providing a formalized description of the LLMs jailbreak attack and defense procedure. Consider a large language model $L$ that generates a response $R_i$ to the user’s prompt $P_i$ at the $i$-th turn, conditioned on the dialogue history $H_{i-1}$, i.e.,
\begin{equation}
R_i = L(P_i, H_{i-1})
\end{equation}

The dialogue history $H_{i-1}$ consists of the system prompt $P_{\text{sys}}$ and the sequence of previous user inputs and model responses up to turn $i - 1$, which means,
\begin{equation}
    H_{i-1} = \{P_{\text{sys}}, P_1, R_1, \dots, P_{i-1}, R_{i-1}\}  
\end{equation}

In the ideal case, we expect the model’s output $R_i$ to be useful, reliable, and safe. For a single-turn attack method $A_{\text{single}}$, the objective is to generate an adversarial prompt $P'_1$ based on a harmful attack goal $G$,
\begin{equation}
    P'_1 = A_{\text{single}}(G)
\end{equation}
When $P'_1$ is submitted to the large language model $L$, it results in a harmful response 
$R'_1 = L (P'_1, H_0)$ 
that aligns with the attack goal $G$, thereby compromising the model’s safety alignment mechanisms.

For a multi-turn attack method $A_{\text{multi}}$, the goal is to construct a sequence of adversarial prompts, referred to as an inductive question chain, denoted by $C = \{P'_1, P'_2, \dots, P'_T\}$, based on a harmful attack goal $G$, i.e., 
\begin{equation}
    C = A_{\text{multi}}(G)
\end{equation}
where $T$ is the maximum number of dialogue turns.
Throughout the multi-turn interaction, this chain is designed to gradually steer the model into generating increasingly unsafe content, ultimately leading to a harmful response $R'_T = L (P'_T, H_{T-1})$ that fulfills the attack goal $G$ and breaches the model’s safety alignment.

When evaluating whether a jailbreak attempt is successful, we typically rely on a judgement function $J$, defined as follows:
\begin{equation}
    J(R_i, G) = 
    \begin{cases}
        1, & \text{if } R_i \text{ satisfies the harmful objective } G \\
        0, & \text{otherwise}
    \end{cases}
\end{equation}
This judgement function serves as a binary indicator, where $J(R_i, G)=1$ denotes a successful jailbreak where the model’s response $R_i$ aligns with the adversary’s harmful intent $G$, and $J(R_i, G) = 0$ indicates that the attack has failed.
The classification function $J$ can be implemented in various forms, including human annotation~\cite{wei2023jailbreak}, prefix matching~\cite{zou2023universal}, LLM-based evaluations~\cite{chao2025jailbreaking}.

For a defense method $D$, it intervenes in the generation process of the language model $L$ through various mechanisms, ultimately producing a defended model denoted as $D * L$. An effective defense enhances the safety of the output response: 
\begin{equation}
    R_i = D * L(P_i, H_{i-1})
\end{equation}

\subsection{Bidirectional Intention Inference Defense}
We propose Bidirectional Intention Inference Defense (BIID) as a defense mechanism against jailbreak attacks targeting LLMs. This approach guides LLMs to dynamically detect latent risks throughout the interaction, thereby establishing a proactive defense framework. 
At its core, BIID integrates forward request-based intention inference with backward response-based intention retrospectio, enabling the model to achieve a semantic-intent joint representation of both user prompts and generated responses. 
This mechanism significantly enhances the model’s ability to identify complex adversarial attempts.

\subsubsection{Forward Request-based Intention Inference}
This forward request-based intention inference mechanism explicitly prompts the model to perform direct intention inference on user inputs, enabling effective identification of latent risks concealed behind seemingly benign requests (before generating responses).

Given a request input $P_i$, we explicitly prompt the model $L'$ to perform intention inference by extracting the underlying intent $I$ and making a binary decision $d$ on whether the request should be refused. This can be formally represented as:
\begin{equation}
    I, d = L'(P_{\text{pre}}(P_i))
\end{equation}
where $P_{\text{pre}}$ denotes the prompt template that instructs the model to perform "Forward Request-based Intention Inference", and $d \in \{0, 1\}$ is a binary decision, where 1 indicates that the request should be refused, and 0 indicates it is acceptable.

While harmful prompts may trigger unsafe outputs when directly submitted to the model, explicitly instructing the model to analyze intent, particularly when requiring it to make an explicit binary decision on “whether the request should be refused”, significantly lowers the activation threshold of its internal safety alignment mechanisms. 
This, in turn, enhances the model’s sensitivity to attack-oriented semantic patterns.
As a result, the model becomes more capable of detecting characteristics of malicious inputs such as semantic obfuscation, semantic association, semantic drift, and goal-oriented manipulation. 
It is then able to proactively reject unsafe requests, effectively blocking attacks at their source before harmful content is ever generated.

\subsubsection{Backward Response-based Intention Inference}
When harmful intent is deeply concealed that the forward intention analysis is evaded by contextual deception strategies (such as disguising prompts as for “educational purposes” or within a “fictional narrative”), the backward intention inference mechanism provides a compensation of risk detection driven by the model’s response.
This mechanism focuses on evaluating the potential harmfulness of the generated response. By performing reverse causal reasoning over the output, it infers the possible user intentions or latent adversarial trajectories that could have led to such a response, thereby reinforcing overall model robustness against sophisticated jailbreak strategies.

Similar to forward intention inference, this backward intention inference mechanism can be formally defined as:
\begin{equation}
    I, d = L'(P_{\text{post}}(R_i))
\end{equation}
where $P_{\text{post}}$ is the prompt template that instructs the model to perform ``Backward Response-based Intention Inference'', and $R_i = L(P_i, H_{i-1})$ is the response generated by the original model $L$ to the user input $P_i$, given the prior dialogue history $H_{i-1}$.
This backward inference mechanism serves as a complementary safeguard, enabling the system to trace harmful intent from response content, particularly in cases where forward analysis is misled by sophisticated prompt obfuscation.

\bigskip
In the face of multi-turn jailbreak attacks characterized by high stealth and logical deception, BIID is capable of deconstructing the latent malicious intent embedded in user queries. 
Even when attackers employ stepwise semantic obfuscation or exploit contextual accumulation to build progressive attack paths, BIID can intercept threats early, before the adversarial reasoning chain is completed. 
This ensures the model’s outputs remain aligned with safety standards and compliant with content policies.

\section{Experiments and Results}
We conduct extensive experiments to evaluate the effectiveness of the proposed BIID approach on multiple LLMs, and compare its performance with existing defense approaches against a variety of jailbreak attacks. 

\begin{table*}[htb]
    \centering
    \setlength{\tabcolsep}{1.5mm}
    \fontsize{9pt}{9.5pt}\selectfont
    \begin{tabular}{c|c|cccccccc|cc}
        \toprule
        \multirow{2}{*}{\textbf{Models}} & \multirow{2}{*}{\textbf{Defense Methods}} & \multicolumn{8}{|c|}{\textbf{Single-Turn Attack}} & \multicolumn{2}{c}{\textbf{Multi-Turn Attack}} \\
        & & AIM & BetterDAN & GCG & ICA & Future & Past & PAIR & RanSearch & Crescendo & Actor \\
        \midrule
        \multicolumn{12}{c}{\texttt{\textbf{Dataset: JailBreakBench}}} \\
        \midrule
        \multirow{9}{*}{\textit{Llama3-8B}} 
            & None & 43.0 & 12.0 & 10.0 & 0.0 & 10.0 & 25.0 & 18.0 & 87.0 & 23.0 & 47.0 \\
            & RPO & 28.0 & 13.0 & 3.0 & 0.0 & 10.0 & 23.0 & 18.0 & 85.0 & 6.0 & 39.0 \\
            & ICD & 31.0 & 11.0 & 5.0 & 0.0 & 3.0 & 11.0 & 18.0 & 81.0 & 15.0 & 38.0 \\
            & Paraphrase & 7.0 & 0.0 & 4.0 & 1.0 & 7.0 & 19.0 & 15.0 & 18.0 & 14.0 & 23.0 \\
            & SelfReminder & 1.0 & 2.0 & 2.0 & 0.0 & 0.0 & 3.0 & 6.0 & 79.0 & 5.0 & 23.0 \\
            & SelfDefense & 13.0 & 0.0 & 3.0 & 0.0 & 2.0 & 7.0 & 6.0 & 25.0 & 6.0 & 24.0 \\
            & SmoothLLM & 58.0 & 17.0 & 4.0 & 0.0 & 3.0 & 12.0 & 9.0 & 52.0 & 15.0 & 40.0 \\
            & SemanticSmooth & 1.0 & 0.0 & 4.0 & 1.0 & 11.0 & 10.0 & 11.0 & 1.0 & 13.0 & 35.0 \\
            & \textbf{BIID(Ours)} & \textbf{0.0} & \textbf{0.0} & \textbf{0.0} & \textbf{0.0} & \textbf{0.0} & \textbf{3.0} & \textbf{1.0} & \textbf{0.0} & \textbf{2.0} & \textbf{0.0} \\
        \midrule
        \multirow{9}{*}{\textit{Llama3-70B}} 
            & None & 51.0 & 90.0 & 6.0 & 21.0 & 7.0 & 18.0 & 16.0 & 46.0 & 13.0 & 35.0 \\
            & RPO & 33.0 & 78.0 & 3.0 & 2.0 & 5.0 & 10.0 & 12.0 & 52.0 & 9.0 & 29.0 \\
            & ICD & 37.0 & 90.0 & 1.0 & 2.0 & 0.0 & 4.0 & 9.0 & 62.0 & 9.0 & 28.0 \\
            & Paraphrase & 6.0 & 0.0 & 1.0 & 0.0 & 8.0 & 20.0 & 12.0 & 21.0 & 11.0 & 27.0 \\
            & SelfReminder & 10.0 & 64.0 & 0.0 & 0.0 & 0.0 & 0.0 & 7.0 & 50.0 & 0.0 & 10.0 \\
            & SelfDefense & 6.0 & 0.0 & 1.0 & 1.0 & 0.0 & 3.0 & 6.0 & 17.0 & 6.0 & 9.0 \\
            & SmoothLLM & 6.0 & 29.0 & 3.0 & 9.0 & 2.0 & 14.0 & 10.0 & 25.0 & 9.0 & 32.0 \\
            & SemanticSmooth & 44.0 & 1.0 & 2.0 & 27.0 & 5.0 & 13.0 & 11.0 & 3.0 & 40.0 & 20.0 \\
            & \textbf{BIID(Ours)} & \textbf{0.0} & \textbf{1.0} & \textbf{1.0} & \textbf{0.0} & \textbf{0.0} & \textbf{3.0} & \textbf{1.0} & \textbf{1.0} & \textbf{0.0} & \textbf{0.0} \\
        \midrule
        \multicolumn{12}{c}{\texttt{\textbf{Dataset: HarmBench}}} \\
        \midrule
        \multirow{9}{*}{\textit{Llama3-8B}} 
            & None & 39.5 & 14.0 & 5.5 & 2.5 & 5.0 & 21.0 & 36.0 & 84.5 & 22.5 & 37.5 \\
            & RPO & 22.0 & 20.0 & 3.0 & 0.0 & 7.5 & 19.5 & 29.5 & 83.0 & 12.0 & 62.0 \\
            & ICD & 22.0 & 10.5 & 3.0 & 0.0 & 1.0 & 13.0 & 35.0 & 72.5 & 12.5 & 52.5 \\
            & Paraphrase & 1.5 & 0.5 & 1.0 & 0.5 & 11.0 & 18.5 & 23.0 & 16.5 & 6.5 & 36.5 \\
            & SelfReminder & 1.5 & 2.0 & 1.0 & 0.5 & 0.0 & 2.0 & 20.0 & 69.0 & 8.0 & 33.5 \\
            & SelfDefense & 11.5 & 0.0 & 4.0 & 0.0 & 4.5 & 7.0 & 16.0 & 36.5 & 7.5 & 26.5 \\
            & SmoothLLM & 43.0 & 20.0 & 7.0 & 0.0 & 1.5 & 8.0 & 21.5 & 39.5 & 9.5 & 35.0 \\
            & SemanticSmooth & 0.5 & 6.5 & 4.5 & 1.0 & 9.5 & 21.5 & 32.5 & 1.5 & 11.5 & 38.5 \\
            & \textbf{BIID(Ours)} & \textbf{0.5} & \textbf{1.0} & \textbf{0.5} & \textbf{1.0} & \textbf{0.0} & \textbf{0.5} & \textbf{3.5} & \textbf{0.0} & \textbf{1.0} & \textbf{0.5} \\
        \midrule
        \multirow{9}{*}{\textit{Llama3-70B}} 
            & None & 40.5 & 92.0 & 8.0 & 16.5 & 11.5 & 22.5 & 38.5 & 45.0 & 14.5 & 49.5 \\
            & RPO & 22.5 & 77.5 & 8.0 & 0.5 & 4.0 & 11.5 & 39.5 & 38.0 & 11.0 & 42.0 \\
            & ICD & 33.0 & 91.0 & 1.0 & 0.5 & 1.0 & 8.5 & 31.0 & 55.5 & 10.0 & 41.0 \\
            & Paraphrase & 4.5 & 0.0 & 0.5 & 2.0 & 8.5 & 12.5 & 22.5 & 15.5 & 11.0 & 31.0 \\
            & SelfReminder & 9.5 & 59.5 & 1.0 & 0.0 & 0.0 & 4.0 & 29.0 & 44.0 & 4.5 & 21.5 \\
            & SelfDefense & 10.0 & 2.0 & 5.5 & 3.0 & 5.0 & 10.0 & 17.5 & 17.5 & 9.0 & 14.5 \\
            & SmoothLLM & 7.5 & 25.0 & 10.5 & 4.5 & 7.5 & 13.0 & 37.0 & 18.0 & 7.0 & 42.0 \\
            & SemanticSmooth & 0.0 & 5.5 & 4.0 & 10.0 & 6.5 & 16.0 & 22.5 & 1.0 & 13.0 & 34.5 \\
            & \textbf{BIID(Ours)} & \textbf{0.5} & \textbf{3.0} & \textbf{0.5} & \textbf{1.0} & \textbf{3.0} & \textbf{1.5} & \textbf{5.0} & \textbf{0.0} & \textbf{0.5} & \textbf{1.0} \\
        \bottomrule
    \end{tabular}
    \caption{
        Attack Success Rates (ASR) of different defense methods against various attack types across LLMs with different scales (8B and 70B). Lower ASR indicates better defense performance.
    }
    \label{asr_tab}
\end{table*}

\begin{table*}[htb]
    \centering
    \setlength{\tabcolsep}{1mm}
    \fontsize{9pt}{11pt}\selectfont
    \begin{tabular}{ccccccccccc}
        \toprule
        \multirow{2}{*}{\textbf{Models}} & \multirow{2}{*}{\textbf{\makecell{Multi-turn \\ Safety Datasets} }} & \multicolumn{9}{c}{\textbf{Defense Methods}} \\
        \cmidrule(rl){3-11}
        & & None & RPO & ICD & Paraphrase & SelfReminder & SelfDefense & SmoothLLM & SemanticSmooth & \textbf{BIID(Ours)} \\
        \midrule
        \multirow{3}{*}{\textit{Llama3-8B}} 
        & MHJ & 57.93 & 54.33 & 45.58 & 23.80 & 23.88 & 14.07 & 46.45 & 38.01 & \textbf{1.39} \\
        & SafeDialBench & 15.56 & 2.63 & 9.46 & 8.51 & 3.97 & 3.38 & 5.80 & 7.84 & \textbf{1.86} \\
        & CoSafe & 4.74  & 4.81 & 5.71 & 7.62 & 0.95 & 3.32 & 3.83 & 6.25 & \textbf{0.47} \\
        \midrule
        \multirow{3}{*}{\textit{Qwen3-8B}} 
        & MHJ & 53.43 & 48.83 & 41.66 & 19.37 & 22.55 & 23.52 & 44.18 & 31.45 & \textbf{0.69} \\
        & SafeDialBench & 3.24  & 1.97 & 2.51 & 2.70 & 0.00 & 0.00 & 2.04 & 1.34 & \textbf{0.00} \\
        & CoSafe & 1.95 & 0.49 & 0.00 & 1.51 & 0.48 & 0.48 & 0.00 & 1.96 & \textbf{0.47} \\
        \midrule
        \multirow{3}{*}{\textit{Llama3-70B}} 
        & MHJ & 42.74 & 36.29 & 25.37 & 12.03 & 11.19 & 7.91  & 22.22 & 33.07 & \textbf{0.00} \\
        & SafeDialBench & 14.68 & 7.09 & 8.49 & 3.24 & 1.28 & 0.00 & 10.06 & 8.97 & \textbf{1.24} \\
        & CoSafe & 5.79 & 1.90 & 1.42 & 2.89 & 0.00 & 1.46 & 1.95 & 2.87 & \textbf{0.00} \\
        \bottomrule
    \end{tabular}
    \caption{
        Attack Success Rate (ASR) of different defense methods on multi-turn safety datasets. Lower ASR indicates better defense performance. 
    }
    \label{dataset_asr_tab}
\end{table*}

\subsection{Experimental Setup}
\subsubsection{Dataset}
To evaluate the defense effectiveness of our proposed method, we conducted experiments on JailBreakBench ~\cite{chao2024jailbreakbench} and HarmBench ~\cite{mazeika2024harmbench}, against various attack methods.
For HarmBench we use the standard subset (n=200).
To further assess the robustness of our approach in multi-turn interaction settings, we additionally tested on the Multi-Turn Human Jailbreaks (MHJ) dataset~\cite{li2024llmnotrobust}, SafeDialBench~\cite{cao2025safedialbench}, and Cosafe~\cite{yu2024cosafe}. 
For MHJ, we use the DERTA subset (n=144). For SafeDialBench and Cosafe, we constructed two balanced sub-datasets by stratified sampling from the original datasets (n=213 and n=167). 
To measure the general performance degradation introduced by our defense mechanism, we employed AlpacaEval~\cite{dubois2024length} as the benchmark for evaluating utility preservation.
Further detail can be found in the Appendix.

\subsubsection{Attack Methods}
For single-turn attack methods, we selected the static jailbreak prompt templates AIM and BetterDAN from \texttt{jailbreakchat.com} as representative baselines. We also evaluated more advanced methods including GCG~\cite{zou2023universal}, RandomSearch~\cite{andriushchenko2024jailbreaking}, PAIR~\cite{chao2025jailbreaking}, In-Context Attack (ICA)~\cite{wei2023jailbreak}, as well as prompt rewriting techniques that modify the tense of queries into past and future forms~\cite{andriushchenko2024does}.
For multi-turn attack methods, we conducted evaluations using Crescendo~\cite{russinovich2024great} and ActorAttack~\cite{ren2024llmsactorattack} to assess the effectiveness of our defense against progressive and interactive jailbreak strategies.

\subsubsection{Defense Baselines}
For baseline comparisons, since our approach belongs to the category of external defenses, we selected a range of representative external defense methods for evaluation. 
These include In-Context Defense (ICD)~\cite{wei2023jailbreak}, Paraphrase~\cite{jain2023baseline} , RPO~\cite{zhou2024robust}, SelfDefense~\cite{phute2023llm}, SelfReminder~\cite{xie2023defending}, SmoothLLM~\cite{robey2023smoothllm}, and SemanticSmoothLLM~\cite{ji2024defending}.

\subsubsection{Models}
We selected \textit{Llama-3.1-8B-Instruct}, \textit{Llama-3.3-70B-Instruct}~\cite{grattafiori2024llama}, and \textit{Qwen3-8B}~\cite{yang2025qwen3} as the target models for evaluation. 
Notably, we employed \textit{GPT-4o-2024-11-20}~\cite{hurst2024gpt} as the judge model, utilizing the prompt template from the PAIR method~\cite{chao2025jailbreaking} to determine whether the outputs of the LLMs constitute successful jailbreaks.

\subsection{Experimental Results}

\subsubsection{Robust Safety Performance against Diverse Attack Methods}

We compared our method with other defense methods in terms of attack success rate (ASR) when facing different attack methods on two commonly used safety evaluation datasets, JailBreakBench and HarmBench. Table \ref{asr_tab} shows the comparison results of experiments on different scale models (8B and 70B).

\begin{figure}[htbp]
    \centering
    \includegraphics[width=0.95\columnwidth]{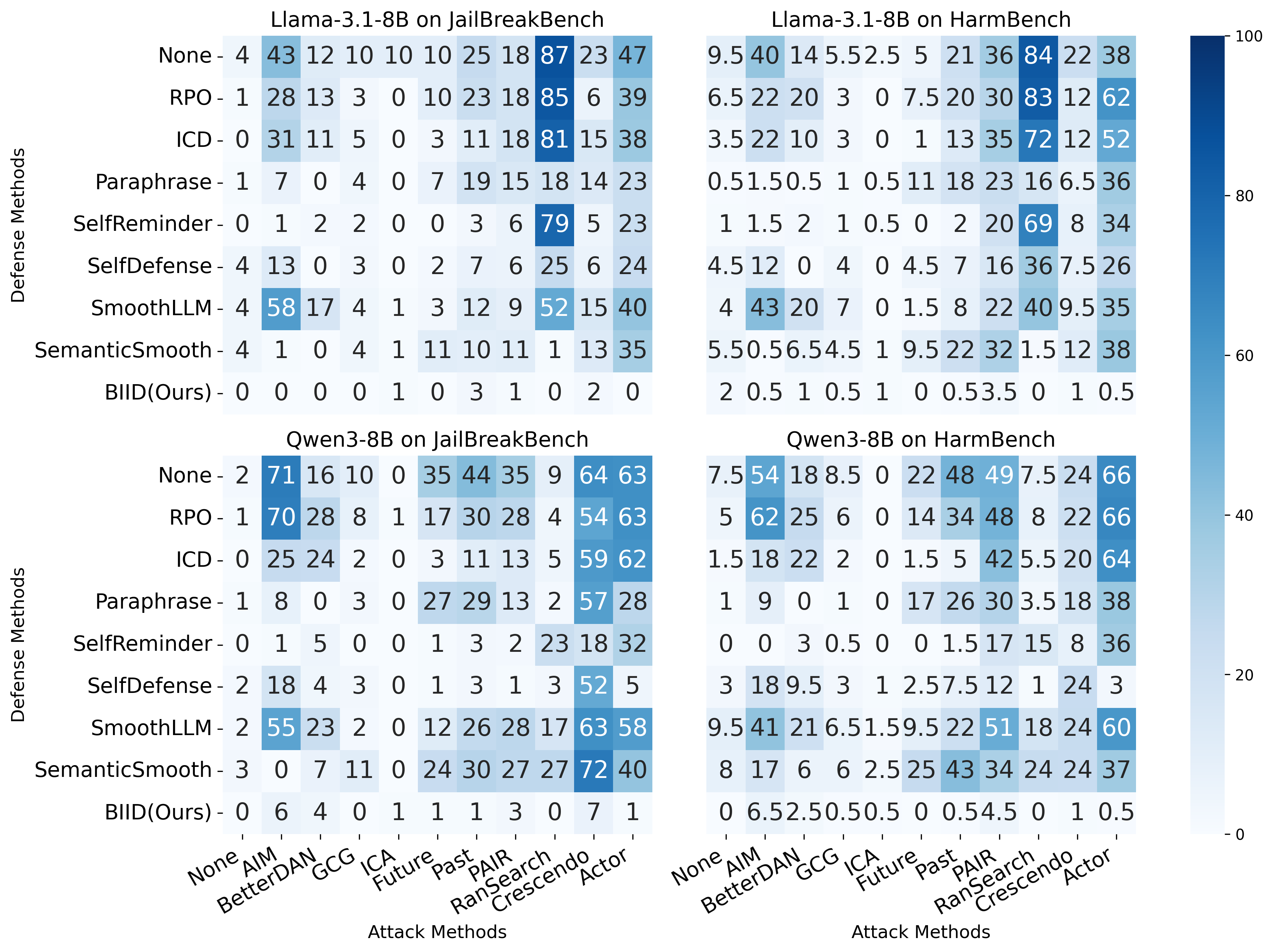}
    \caption{
        Heatmap of ASR of different defense methods against various attack strategies on models of different architecture (\textit{Llama} and \textit{Qwen}).
    }
    \label{qwen3_heatmap_fig}
\end{figure}


From the table we observed that, our method reduces ASR to near or equal to 0\% against all attack methods across different datasets and target models, indicating the best safety performance. 
Compared to other methods, our approach demonstrates a clear advantage in both multi-turn and single-turn attack scenarios.
Especially in multi-turn scenarios, other defense methods perform poorly. Particularly, under the Actor attack on the 8B model, other defense methods can only reduce ASR to about 20\% , while BIID can reduce ASR to nearly 0\%. This demonstrates the superiority of our method over other methods when facing multi-turn attacks.
Besides, other defense methods lack consistent performance across diverse attack scenarios, showing effectiveness against certain attacks while performing poorly against others.
For example, SmoothLLM fails to defend against the AIM attack on the 8B model (ASR = 58\% on JailBreakBench, 43\% on HarmBench), while SelfReminder completely breaks down against the BetterDAN attack on the 70B model (ASR = 64\% on JailBreakBench, 59.5\% on HarmBench).
We also observed that other defense methods exhibited fluctuations in safety performance across models of different scales. For instance, SemanticSmoothLLM reduces the ASR to 1\% against the ICA attack on the 8B model, but its ASR rises sharply to 27\% on the 70B model.
By contrast, our method demonstrates overall robustness across different attack methods and models.

To further investigate the impact of architectural differences, we conduct a comparative analysis between \textit{Llama-3.1-8B-Instruct} and \textit{Qwen3-8B}. 
As shown in the figure \ref{qwen3_heatmap_fig}, our approach achieves the best average performance across both model architectures, demonstrating strong cross-model consistency.
We also observed that some defense methods performed unevenly across different architecture models. SemanticSmoothLLM is effective against the RandomSearch attack on \textit{Llama-3.1} (ASR=1\% on JailBreakBench, 1.5\% on HarmBench) but shows significantly reduced effectiveness on \textit{Qwen3} (ASR=27\% on JailBreakBench, 24\% on HarmBench).
Additionally, we find that \textit{Qwen3} generally exhibits higher ASR compared to \textit{Llama-3.1}. 
This discrepancy may stem from \textit{Qwen3}'s use of chain-of-thought (CoT) fine-tuning, which prompts the model to explicitly generate intermediate reasoning steps. These intermediate steps may inadvertently surface unsafe content, contributing to an elevated overall ASR.

\subsubsection{Efficient Defense on Multi-turn Safety Datasets}

To further evaluate the performance of different defense strategies under multi-turn attack scenarios, we conducted experiments comparing BIID with none-defense baseline and other defense methods on the multi-turn safery datasets MHJ, SafeDialBench, and CoSafe, as shown in Table \ref{dataset_asr_tab}. 
The ASR of the no-defense baseline on each dataset can be approximately regarded as an indicator of the defense difficulty associated with that dataset.
On the most challenging dataset MHJ (ASR$>$40\%), other defense methods can only partially reduce the ASR, with the lowest ASRs across the three models being 14.07\%, 19.37\% and 7.91\% respectively. In contrast, BIID can reduce the ASR to nearly 0\%, which demonstrates a substantial performance advantage over all other defense methods.
Meanwhile, BIID demonstrates strong performance on the other two datasets as well, reducing the ASR on \textit{LLama-3.1-8B} to 1\% and 2\% seperately. In general, our method has achieved the best overall effect on the three datasets.

On the other hand, we observe that the effectiveness of other defense methods tends to degrade as the capability of the target model decreases. Specifically, comparing performance on the MHJ dataset between \textit{Llama3-70B} and \textit{Llama-8B}, the ASR of the no-defense baseline increases by approximately 15\% on the smaller model. 
For other defense methods, ICD sees an increase of about 20\%, and SmoothLLM about 24\%. 
In contrast, our method exhibits an increase of less than 2\%,  which is the smallest among all methods. 
This indicates that the performance of our defense is minimally affected by the underlying model’s capacity, providing nearly equivalent protection for smaller models as it does for larger ones.

\subsubsection{Optimal Balance between General Utility and Defense Effectiveness}

\begin{figure}[t]
    \centering
    \includegraphics[width=0.95\columnwidth]{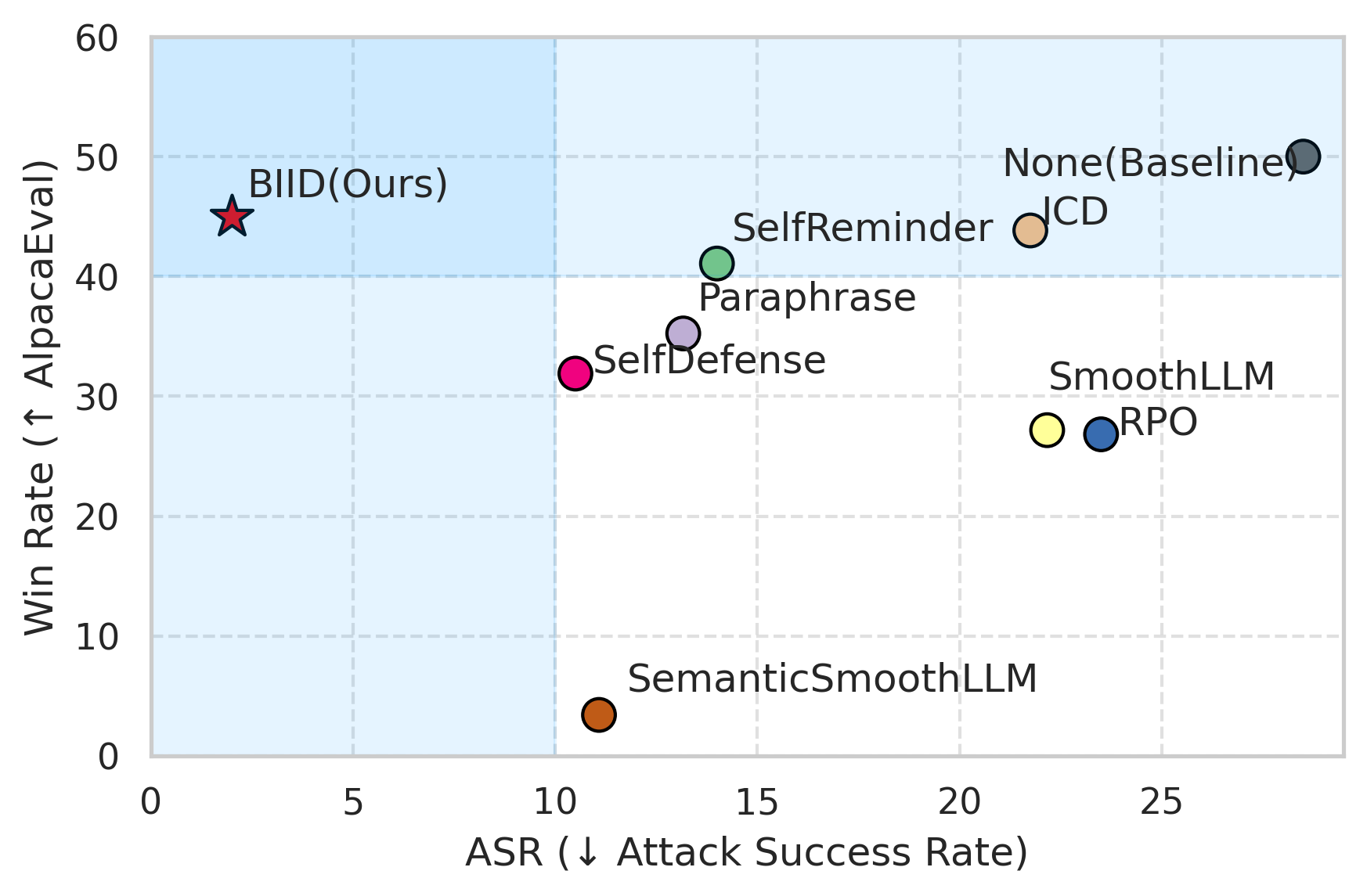}
    \caption{
        Trade-off between defense effectiveness and general performance on  \textit{Llama-3.1-8B-Instruct}. 
        Methods located in the upper-left area demonstrate superior performance by achieving the best trade-off between safety (↓ ASR) and utility preservation (↑ Win Rate).
    }
    \label{asr_winrate_fig}
\end{figure}

To quantify the impact of different defense methods on the general capabilities of LLMs, we conducted a systematic evaluation using the AlpacaEval benchmark. All win rates were computed relative to the outputs of the original, unprotected model. Figure \ref{asr_winrate_fig} illustrates the trade-off between defense effectiveness and general performance on \textit{Llama-3.1-8B-Instruct model}. From Figure~\ref{asr_winrate_fig}, methods located in the upper-right region (such as SelfReminder, ICD) preserve the model’s general utility well but exhibit limited defense capability; those near the lower-left (e.g., SelfDefense, SemanticSmoothLLM) tend to achieve stronger defense but at the cost of significant utility degradation. 

Our method appears in the upper-left region of the figure, indicating that it not only provides strong defense performance but also maintains a high level of general-purpose capability. This demonstrates that our approach achieves a favorable balance between robustness and utility.
This advantage arises from the fact that our method operates as a plug-and-play external module, maintaining strong generality and flexibility without modifying the original model’s inputs or outputs. As a result, it does not interfere with the model’s generation quality and preserves the original model’s general-purpose capabilities effectively.

\subsubsection{Intention Detection Phase Analysis}
\begin{figure}[ht]
\centering
\includegraphics[width=0.95\columnwidth]{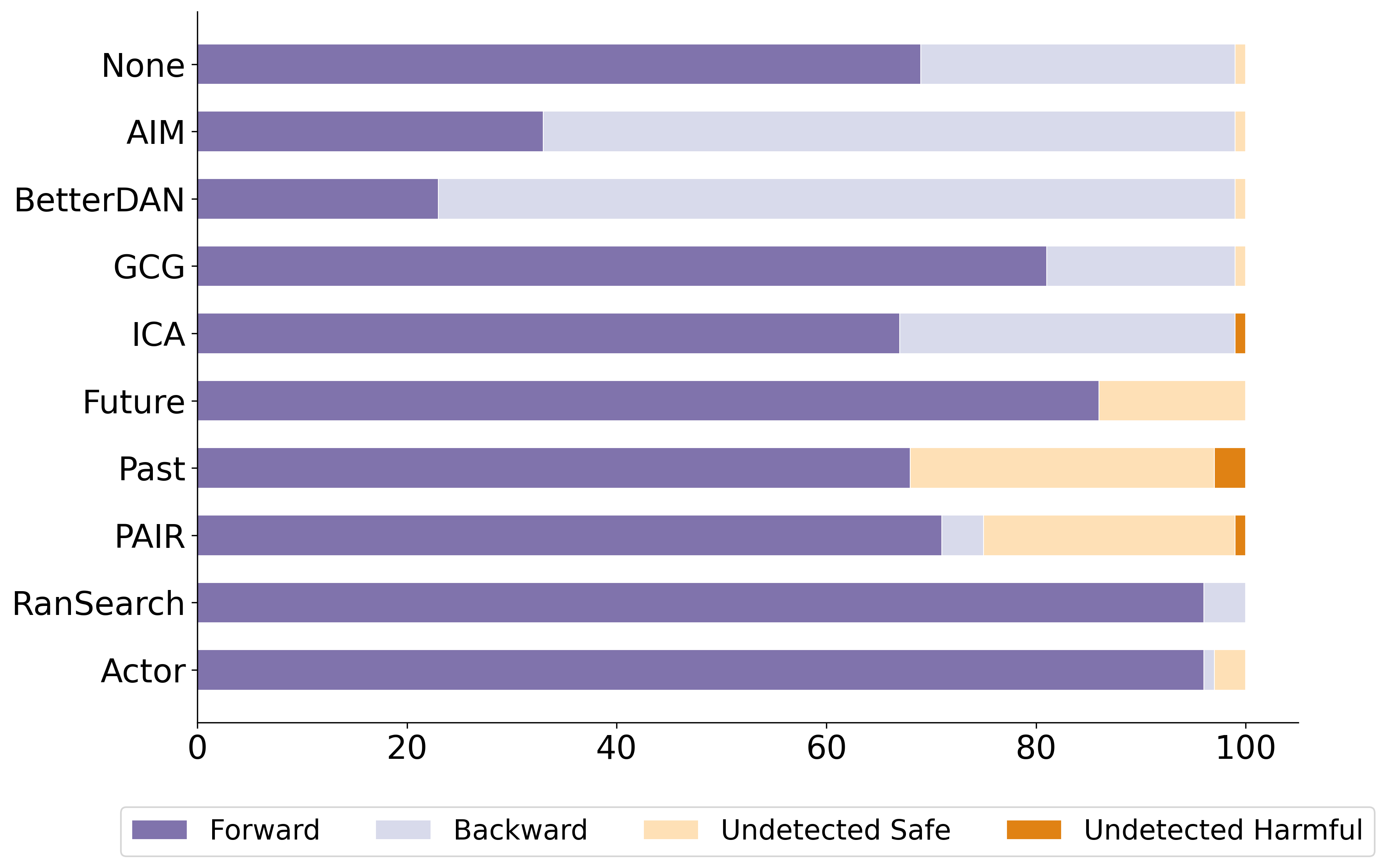}
\caption{
    Phase distribution of malicious intent detection by BIID across varying attack methods.
}
\label{phase_ana_fig}
\end{figure}

We further perform a intention detection phase analysis of our BIID method to examine at which stage each the malicious intention of a quest is successfully detected and intercepted. 
This analysis aims to evaluate BIID’s ability to detect various adversarial intents at different stages of interaction.
Specifically, we categorize whether the adversarial intention in a given prompt is:
detected during the forward intention inference phase,
inferred during the backward intention retrospection phase, or
not detected by either inference phase, as shown in Figure \ref{phase_ana_fig}.

Based on the heterogeneity of attack strategies, our defense method exhibits three distinct behavioral response patterns:
First, forward-dominant detection (e.g., None, GCG, ICA, RandomSearch, ActorAttack):
In this category, the majority of harmful requests are successfully intercepted via forward intention inference. 
Specifically, white-box optimization-based attacks (GCG, RandomSearch) are rendered ineffective due to enforced intention parsing, which neutralizes their adversarial suffixes. 
Context-accumulating attacks (ICA, ActorAttack) fail to build up malicious chains, enabling forward mechanisms to easily detect the underlying intent.
Second, backward-dominant detection (e.g., AIM, BetterDAN):
Here, harmful intents largely bypass forward intention filtering by exploiting role-playing and persona-based strategies, but are subsequently exposed through backward intention analysis. The generated responses reveal latent risk signals that betray the hidden malicious intent, which is effectively captured via backward causal reasoning.
Third, residual-safe acceptance (e.g., Future, Past, PAIR, Crescendo):
In these cases, many adversarial attempts are not explicitly rejected by either forward or backward modules. However, the final outputs remain safe. 
This is attributed to the attackers' efforts to evade alignment constraints, which often result in diluted semantic aggression. Consequently, the generated content diverges from the original harmful intent, reducing the actual risk.

The above analysis summarizes the distinct behavioral patterns exhibited by BIID when confronted with attack methods of varying characteristics. These observations help explain why BIID can effectively defend against diverse attack strategies, highlighting its robustness across different types of adversarial methods. 
This underscores the resilience of our approach in complex and varied attack scenarios.

\section{Conclusion}
This study proposes an effective bidirectional intention inference defense method against the multi-turn jailbreak attacks. Through extensive experiments across three LLMs on JailBreakBench and HarmBench, covering eight single-turn and two multi-turn jailbreak attack methods, as well as evaluations on multi-turn safety datasets including MHJ, SafeDialBench, and CoSafe, our method consistently achieves the lowest attack success rate compared to both the no-defense baseline and seven existing defense methods, demonstrating superior safety performance.
Notably, our approach shows significant advantages in multi-turn scenarios, where existing defenses typically struggle. 
Furthermore, results on AlpacaEval indicate that our method ensures optimal safety performance while preserveing the model’s general capabilities to a high degree.
Overall, this work presents an initial yet meaningful exploration into enhancing the output safety of large language models in multi-turn interactions, contributing toward the trustworthy deployment of LLMs in real-world applications.
In the future, intention inference based methods could be explored to endow LLMs with dynamic adaptability during interactions, as well as the ability to construct user profiles. This approach has the potential to enhance both the safety of LLMs and the quality of their generated content.

\bibliography{ref}


\end{document}